\documentclass[aps,prl,twocolumn,showpacs]{revtex4}  
\usepackage{bm}
\usepackage{graphicx}
\usepackage{amsmath}
\usepackage{amssymb}
\usepackage[dvips]{color}
\usepackage{subfigure}

\begin{document}
\title{Non-monotonic dependence of the friction coefficient on heterogeneous stiffness}
\author{F. Giacco$^{1,2}$, L. Saggese$^{3}$, L. de Arcangelis$^{3,4}$,   M. Pica Ciamarra$^{5,1}$, E. 
Lippiello$^{2,4}$}
\affiliation{$^{1}$CNR--SPIN, Dept. of Physics, University of Naples ``Federico II'', 
Naples, 
Italy\\
$^{2}$Dept. of Mathematics and Physics, Second University of Naples and CNISM, 
Caserta, Italy \\
$^{3}$Dept. of Industrial and Information Engineering, Second University of 
Naples and CNISM, Aversa (CE), Italy\\
$^{4}$Kavli Institute for Theoretical Physics, University of 
California, Santa Barbara, CA 93106-4030 USA\\
$^{5}$Division of Physics and Applied Physics, School of
Physical and Mathematical Sciences, Nanyang Technological University, Singapore 
}

\begin{abstract}
The complexity of the frictional dynamics at the microscopic scale makes difficult
to identify all of its controlling parameters.
Indeed, experiments on sheared elastic bodies have shown that the static friction coefficient depends on 
 loading conditions, the real area of contact along the interfaces and the confining pressure.
Here we show, by means of numerical simulations of a 2D Burridge-Knopoff model with a simple local
friction law, that the macroscopic friction coefficient depends non-monotonically on the bulk elasticity of the system.
This occurs because  elastic constants control the geometrical features of the rupture fronts during the stick-slip
dynamics, leading to four different ordering regimes characterized by different
orientations of the rupture fronts with respect to the external shear 
direction. We rationalize these results by means of an energetic balance argument.
\end{abstract}

\maketitle

Frictional forces between sliding objects convert kinetic energy into
heat, and act in systems whose size ranges from the nanometer scale, as in some
micro and nanomachines, up to the kilometer scale, typical of geophysical
processes. The microscopic origin of frictional forces
is therefore deeply investigated,
and strategies to tune their effects are actively sought
~\cite{dowson,baumberger,capozzaprl,rmpvanossi,capozza}. 
In the classical description of frictional processes~\cite{bowden}, the transition from the static to the sliding regime 
occurs as the applied shear stress overcomes the product
of the normal applied force and   the friction coefficient $\mu$ (Amontons--Coulomb law).
However, experiments conducted in the last decade ~\cite{fineb2007,fineb2010} 
have shown that this transition is driven by a local dynamics of frictional interfaces,
which occurs well before macroscopic sliding. 
In addition, studies on the systematic violation of
Amontons--Coulomb law and the dependence on the loading conditions have clarified that the static friction coefficient is not a material
constant ~\cite{SRotsuki2013,fineb2011}.
This is consistent with numerical studies based of 1D~\cite{Maegawa,Braun,tromborg2012}
spring-block models that have clarified the influence of the loading conditions
on the  nucleation fronts. 
In particular they have shown that the friction coefficient
decreases with the confining pressure and the system size.
While the effect of the elasticity of the slider in the direction perpendicular
do the driving one has been recently addressed via the
study of 2D spring-block models \cite{tromborg2011},
the role of the elasticity of the contact surface has not yet been clarified.\\

\begin{figure}
\includegraphics*[scale=0.43]{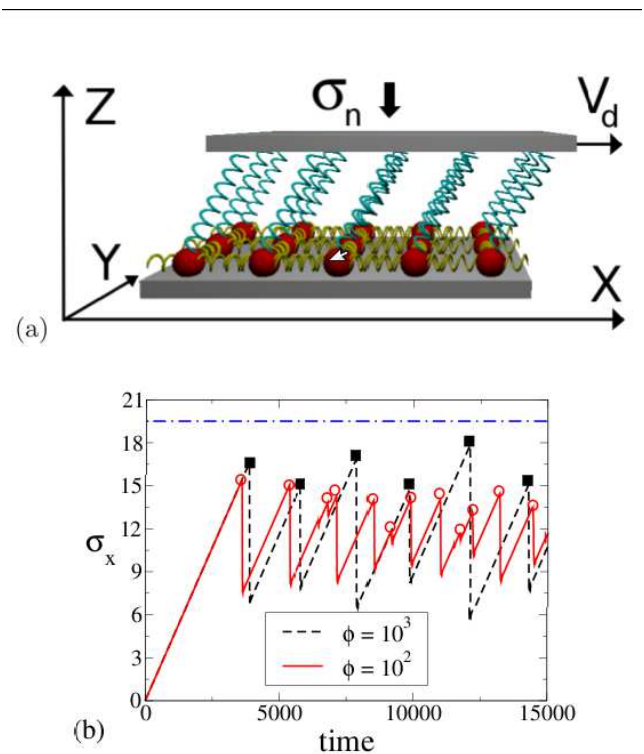}
\caption{The 2D Burridge-Knopoff model.
(a) Schematic representation of the model: an elastic layer of red
particles connected by yellow springs is in contact with a bottom-flat substrate (gray). 
The system is driven by an external spring mechanism along the $x$-direction at constant
velocity $V_{d}$ and each particle is confined by a constant pressure $\textbf{$\sigma_n$}$.
(b) Time dependence of the shear stress $\sigma_{x}(t)$ during the stick-slip
dynamics for two systems with different values of $\phi$. Filled black squares
and open red circles  indicate the values of the shear stress that, divided by
$\textbf{$\sigma_n$}$, are used to estimate the macroscopic friction coefficient
$\overline{\mu}$. The transverse dash-dotted (blue) line indicates  the value of the stress
corresponding to the Amontons-Coulomb threshold.
\label{fig:model}}
\end{figure}

In this study we show that the elasticity of the contact surface influences the features of the fracture fronts
leading to a non-monotonic behaviour of the friction coefficient.
These results are obtained via  numerical simulations of a 2D ($xy$)
spring-block model (Fig.~\ref{fig:model}a), and are supported by analytical arguments. 
Our model, fully described in the method section, is a simple variation
of the Burridge--Knopoff~\cite{bk1,carlson} (BK) model
that is commonly used in seismology to describe a seismic fault under tectonic drive,
and that reproduces many statistical features of earthquake occurrence~\cite{bk2D,bk2Dlr,prl2010,epl2011}. 
We will study its properties as a function of the parameter $\phi=k_{b}/\Delta k_{d}$,
where $k_b$ is the elastic constant of the springs connecting close blocks,
and $\Delta k_{d}$ the variance of the distribution of the 
elastic constant $k_d$ with which each block is coupled to the drive
(see Fig.~\ref{fig:model}a). Accordingly, $\phi$ measures the relevance
of the elastic heterogeneity, the $\phi \to 0$ limit corresponding to a system elastically homogeneous.
We show that, even if the Amontons--Coulomb law is locally satisfied for each contact,
violations at the macroscopic scale are observed, due to the interplay between the elasticity
of the material and the heterogeneity of the local frictional forces.
This interplay influences the macroscopic friction coefficient as it determines
the ordering properties of the system change under shear: stiff systems keep their ordered structure, 
while soft systems disorder more easily. 
We report four different shear--induced ordering regimes, 
each one characterized by rupture fronts with specific geometric features.
The macroscopic friction coefficient $\overline{\mu}$ does not vary monotonically
with the degree of order of the system or with the degree of elastic heterogeneity: Indeed $\overline{\mu}$
exhibits a minimum when the periodic order of the system
is broken in the direction perpendicular to the shear, which occurs at intermediate heterogeneities.
This leads to a $\overline{\mu}$ that is larger both for highly ordered (homogeneous) and highly disordered (heterogeneous) systems.

\begin{figure}
\includegraphics*[scale=0.4]{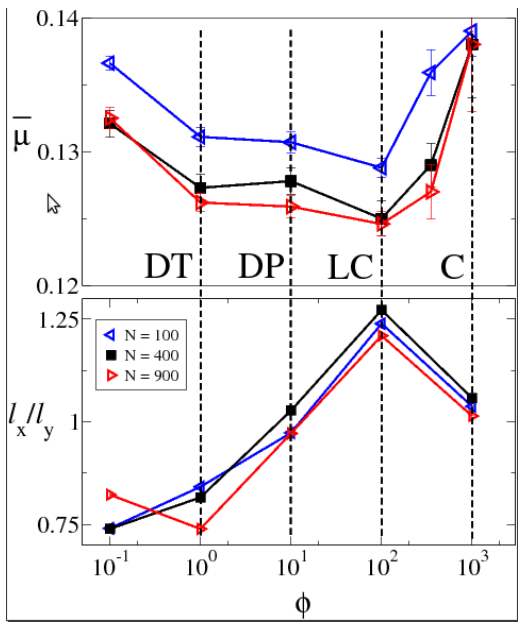}
\caption{The friction coefficion depends non-monotonically on the degree of elastic heterogeneity. (a) the macroscopic friction coefficient as a
function of the parameter $\phi$. The value of  $\overline{\mu}$  is defined as
$\langle \sigma_{\rm fail} \rangle / \sigma_n$, where $\sigma_{\rm fail}$  is the maximum shear stress right before large stress drops (symbols in Fig. 1b) and 
$\sigma_n$ is the confining pressure.
Different symbols refer to different system sizes. The dashed vertical lines indicate four regimes: {\it crystalline} C, {\it laminar crystalline} LC, {\it disorder-parallel} DP and {\it disorder-transverse} DT.  Bottom panel: the asymmetry
factor of the clusters of slipping particles as a function of the parameter
$\phi$.
\label{fig:friction}
}
\end{figure}

The dynamics exhibits the typical stick-slip behaviour with phases of
slow stress accumulation interrupted by an abrupt energy release.
Fig.~\ref{fig:model}b shows for  different values of $\phi$ the time
evolution of the shear stress $\sigma_x(t)= \sum_i^N k_d^i(x_i(t)-x_{i}(0)-V_d t)/L_xL_y$, where $x_i(t)$ is the $x$-coordinate of the $i$-th particle at time $t$. 
The stress drop amplitude exhibits a power law distribution
\cite{prl2010,epl2011} that can be related to the GR law of experimental
seismicity. 
We define the macroscopic friction coefficient as the average over many slips of the steady state shear stress right before
failure (symbols in Fig.~\ref{fig:model}b), divided by the confining pressure $\overline{\mu} = \langle \sigma_{\rm fail} \rangle / \sigma_n $.
The dependence of $\overline{\mu}$ on $\phi$ in Fig.~\ref{fig:friction} is clearly non--monotonic, with a minimum
corresponding to a $\sim 40/\%$ reduction of the friction coefficient with respect to the microscopic value.
This minimum is observed for all values of $N$, and becomes more pronounced for larger $N$.

\begin{figure}
\includegraphics*[scale=0.27]{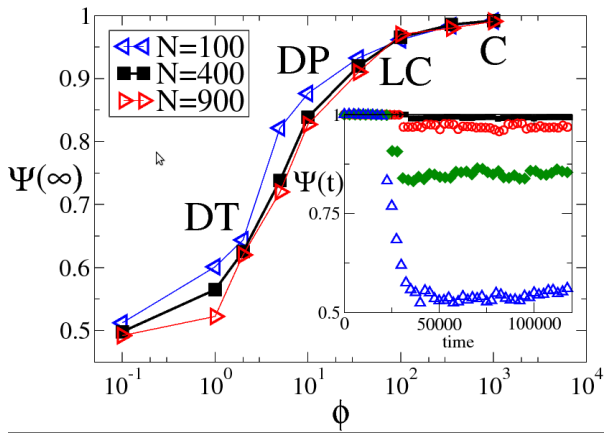}
\caption{The degree of elastic heterogeneities controls the ordering properties of the system. Main panel: the asymptotic value of $\Psi(t)$ as a
function of  the parameter $\phi$. Systems are initially
 prepared in the same ordered state. Depending on  $\phi$, structural changes can occur as revealed by the order
parameter, which  drops from 1 (rigid bond) to $\sim 0.5$ (very elastic bond). Error bars are comparable to symbol sizes.
Inset: Time evolution of $\Psi(t)$ for systems with $N=400$ and $\phi=10^3$ (black filled
squares), $\phi=10^2$ (open red circles), $\phi=10$ (green diamonds) and
$\phi=1$ (blue triangles).
\label{fig:porder}
}
\end{figure}

Next, we show that the minimum of  $\overline{\mu}$ is related to changes of the ordering properties of the elastic surface,
and, to this end, we consider the $\phi$ dependence of the bond--orientation ordering parameter~\cite{nelson}. This is defined as
\begin{eqnarray}
\Psi(t)=\frac{1}{N}\sum_{i=1}^{N}\frac{1}{4}\left\vert\sum_{j=1}^{n_{j}}
e^{
4i\theta_{ij}(t)}\right|,
\label{eqnarray:porder}
\end{eqnarray}
where the second sum runs over all $n_j$ nearest neighbors of particle $i$ and
$\theta_{ij}$ measures the orientation of each bond at time $t$ with respect to the shear direction. 
When the configuration preserves its original ordered square lattice
configuration $\Psi=1$, whereas in the opposite limit of a fully disordered configuration $\Psi\simeq
0.5$.  
Fig.~\ref{fig:porder} (inset) shows that as the systems is sheared $\Psi(t)$
decreases from $\Psi(0) = 1$ to a limiting value $\Psi(\infty)$. 
 The main panel shows that this asymptotic value is a continuously decreasing function of $\phi$, that approaches
its ordered and disordered limits for large and small $\phi$, respectively.
This behavior is consistently observed for different system sizes $N$.

Concerning the dependence of $\Psi$ on the system size $N$ we observe that for $\phi\ge 10^2$ the parameter $\Psi(\infty)$ is $N$ independent.
Conversely, at smaller $\phi$ we notice $\Psi(\infty)$ is a weak decreasing function of $N$, indicating  that the larger is $N$ the more disordered is the configuration.
This behaviour can be attributed to heterogeneities of the local stress. Indeed, for larger systems  the probability  to find local instabilities is higher which favors the occurrence of local rearrangements leading to more disordered configurations. The same argument can be also used to explain the weak decrease of the macroscopic friction with $N$ (Fig. 2a), but does not account for the presence
of the minimum in $\overline{\mu}$. Indeed, $\overline{\mu}$ is not a monotonic function  of $\Psi$. \\

\begin{figure}
\includegraphics*[scale=0.33]{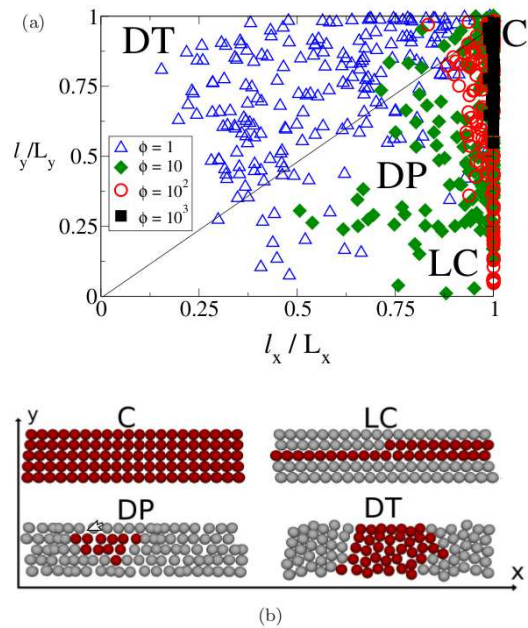}
\caption{The morphology of the clusters of slipping particles depends on the elastic heterogeneitiy. (a) Scatter plot of the clusters dimensions, along the directions 
 parallel and transverse to the shear, for system having different values of $\phi$.
By reducing the stiffness of the system we  observe regions with  different
cluster shapes, indicated by the letters C, LC, DP, DT. 
(b) Schematic representations of the geometry of the clusters   corresponding to the  regions reported in the scatter
plot.  Here we show a system of dimension $L_{x}=20$ and $L_{y}=5$, the same
behaviour is also observed for larger systems.
\label{fig:scatter}
}
\end{figure}

Here we show that $\overline{\mu}$ variations can be related to the geometrical properties of clusters of slipping particles.
We define as ``slipping particles''  those particles with a displacement in the shear direction larger than a given threshold $S_x$. 
We set $S_x=0.01\, l$ and find that particles form compact clusters. 
The geometrical features of the slipping clusters are determined by their sizes along the direction 
parallel $l_x$ and transverse $l_y$ to the shear,   
$l_{x}\equiv B^{-1}\sum_{i,j}\Delta x_{ij}$ and $l_{y}\equiv
B^{-1}\sum_{i,j}\Delta y_{ij}$. Here,  $\Delta x_{ij}$ ($\Delta y_{ij}$) is the distance between particles $i$ and $j$ along the  $x$ $(y)$ direction and the sum is extended  to all $N_c (N_c-1)$ particle couples belonging to a cluster.  The normalization factor $B=N_c (N_c-1)/4$ insures that for a cluster involving the whole system $l_x=L_x$ and $l_y=L_y$.\\
In Fig.~\ref{fig:scatter}a we present a parametric plot of $l_y/L_y$ vs
$l_x/L_x$ for four different values of $\phi$. Cluster configurations can be distinguished into four
classes, determined by $\phi$, whose typical shape is reported  in Fig.~\ref{fig:scatter}b. 
For $\phi \gg 1$ we have the {\it crystalline}  regime (C, black circles), represented by symbols with $l_x \simeq L_x$ and 
In this case, the system behaves as a rigid body and slips involve all particles, keeping the original {\it crystalline} order. 
For smaller value of $\phi$ we have the {\it laminar crystalline} regime (LC, red
squares) where $l_x \simeq L_x$ and $l_y$ with  values in the range
$[1,L_y]$. In this regime, a typical slip involves the motion of one or few
parallel lines. This is consistent with the ordering features observed in this regime (LC lower panel),
characterized by order along the direction of the shear and disorder along the transverse direction.
A further reduction of $\phi$ first breaks order in the shearing direction, giving rise to  the {\it disorder--parallel} regime  (DP, green
diamonds), and then fully disorders the system, giving rise to the  {\it disorder--transverse} regime (DT, blue  triangles).
 The shape of the clusters in both these 
regions are asymmetric with $l_x/L_x>l_y/L_y$ in the DP whereas
$l_x/L_x<l_y/L_y$ in the DT. This information can be directly extracted from
Fig.~\ref{fig:scatter}a where we observe that the DP and the DT regimes
respectively populate regions above and below the diagonal.
We characterize the asymmetry of the cluster shape comparing their average longitudinal
and transverse sizes, $ l_x / l_y $. As shown
in  Fig.~\ref{fig:friction}b, this ratio varies non--monotonically with $\phi$,
and has a maximum  corresponding to the  minimum of $\overline{\mu}$.
This suggests that the lowest value of $\overline{\mu}$ is
obtained when slips involve the  horizontal displacement of the smallest  number of lines.

An explanation of this result and the presence of different regimes is given by a simple energetic argument.
Let us suppose that at given time an amount of energy $E_R$ provided by
the external drive is relaxed via a slip of length $\delta$ such that 
the relaxed energy is $E_R \sim k_d \delta^2$.
This energy can be released through the motion of a rectangular cluster of particles of size $n_x
l\times n_y l$. Assuming that all particles in the cluster rigidly slip of the same distance, the amount of energy released in the slip comes only from the perimeter particles  and is given by
\begin{eqnarray}
\Delta E = k_b n_y \delta^2 \Theta(L_x-n_x l)  + \nonumber\\ + k_b n_x 
\left ((\delta^2+l^2)^{1/2}-l\right )^2 \Theta(L_y-n_y l),
\label{energdelta}
\end{eqnarray}
where the Heaviside theta function takes into account that, because of periodic
boundaries, if a side of the cluster becomes as large as the system size, the
interface contribution vanishes.  
Eq. \ref{energdelta}, in the limit of small slips $\delta \ll l$, becomes 
\begin{eqnarray}
\Delta E \simeq k_b n_y \delta^2 \Theta(L_x-n_x l)+\nonumber\\ + \frac{1}{4}k_b n_x \delta^2 \left (\frac{\delta}{l}\right )^2 \Theta(L_y-n_y l).
\label{energdeltasmall}
\end{eqnarray}
As a consequence, for a rigid system ($k_b\gg k_d$) in order to 
have $\Delta E \sim E_{R}$ the first term in Eq.(\ref{energdeltasmall}) 
must be zero ($n_x l=L_x)$  and also the condition $\delta \ll l$ must be satisfied so that $ k_b n_x \left (\frac{\delta}{l}\right )^2 \sim k_d$.  
This corresponds to the slip of entire rows (C regime in Fig.
\ref{fig:scatter}a). 
However, these slips are possible only if the configuration is ordered. 
When $k_b$ becomes smaller, as indicated by the behavior of $\Phi$ (Fig.3), fluctuations appear in the 
lattice organization preventing slips in the form of entire rows. 
In this case, since  $\delta < l$, configurations with $n_y <n_x$ are
still energetically favored. This situation corresponds to the LC regime in Fig.
\ref{fig:scatter}a. 
On the other hand, as soon as $k_b$ (and $\phi$) becomes sufficiently small so that 
$n_y l=L_y$, the second term  in Eq. \ref{energdeltasmall} vanishes and 
the configurations
corresponding to the DT regime in Fig. \ref{fig:scatter}a
 are  energetically favored. 
Summarizing, slipping clusters for decreasing $k_b$ crossover from 
configurations elongated in the drive direction 
to configuration involving more and more particles in the 
direction perpendicular to the drive. 
Obviously, the transition from ordered to disordered configurations also depends on the degree of heterogeneity of the local shear stress: for larger values of $\Delta k_d$ the transition is expected at larger values of $k_b$.
Moreover, if one changes the shape of the system with $L_x \le L_y$,
the DT regime is never observed. The last statement has been verified
for systems with different  values of the  ratio $L_x/L_y$.\\

In conclusion, we have found that frictional properties 
depend on the geometry of the rupture fronts,
which are determined by the ordering features of the contact interface.
These ordering features are the result of the interface
deformation occurring during the shearing process,
and thus depend upon the elasticity of the material.
Our study  provides a direct relation between bulk elastic properties
and the friction coefficient. \\

\subsection{Methods}
We consider a  two dimensional BK model, where a layer of
particles of mass $m$ is placed on a square lattice with lattice constant $l$, and nearest
neighbor particles are connected by harmonic elastic springs with constant $k_b$ (Fig.~\ref{fig:model}a).
Each particle $i$ is connected to a plate moving with constant velocity
along the $x$ axis by a spring whose stiffness $k_{d}^{i}$ 
is uniformly distributed in the range $(k_{d} - \Delta k_{d},k_{d}+\Delta k_{d})$.
$\Delta k_{d}$ is a parameter allowing to control
the heterogeneity of the local shear stress.
A granular--like approach \cite{silbert} is used to model the interaction of a 
particle with the bottom plate. At time $t$ the frictional force acting on
a particle is given by $\vec F_{s}=k_e \vec {\Delta r}$,
where $\vec {\Delta r} = \int^t_{t_0} \vec v dt$
is the shear displacement of the particle due to creep motion, and $t_0$ the time 
of contact formation. Indeed, each contact breaks and reforms as soon as
the Amontons--Coulomb threshold criterion $|\vec F_{s}|\leq \mu_s\textbf{$\sigma_n$}A$ is violated.
Here 
$\textbf{$\sigma_n$}$  
is the confining normal force acting on each grain, $A=l^2$ the lattice cell area, and $\mu_s$ the local 
coefficient of static friction. The grain motion  is also damped by a viscous term $-m\gamma \vec v$. 
Mass, spring constants and  lengths are expressed in units of $m$, $k_{d}$ and $l$, respectively.
We  fix $F_N=5\, k_{d} l$, $\mu_s=0.2$, $k_e = 10\,k_{d}$, $\Delta k_{d} = k_{d}$, $\sigma_n = 5 \,k_{d}/l$, 
$V_d= 2\cdot10^{-2}\, (m/k_d)^{1/2}$ and $\gamma = 0.2\, (k_d/m)^{1/2}$.
These values insure that simulations are in the quasi-static regime.
Periodic boundary conditions are considered in both directions.
The number of particles $N$ equals the system size $L_x \times L_y$, with $L_y=L_{x}/4$  and  assumes
 the following values, $N = 100,400,900$.
We have investigated the frictional properties of the system as a function of the 
parameter $\phi=k_{b}/\Delta k_{d}$ that we vary by changing $k_b$. 
This parameter measures the relevance of the stiffness of the system
with respect to the heterogeneity of the shearing forces.

\section{Acknowledgements}
We acknowledge the financial support of MIUR–FIRB
RBFR081IUK (2008) and MIUR–PRIN 20098ZPTW7
(2009).

\bibliography{biblio}{}

\begin{thebibliography}{22}
\expandafter\ifx\csname natexlab\endcsname\relax\def\natexlab#1{#1}\fi
\expandafter\ifx\csname bibnamefont\endcsname\relax
  \def\bibnamefont#1{#1}\fi
\expandafter\ifx\csname bibfnamefont\endcsname\relax
  \def\bibfnamefont#1{#1}\fi
\expandafter\ifx\csname citenamefont\endcsname\relax
  \def\citenamefont#1{#1}\fi
\expandafter\ifx\csname url\endcsname\relax
  \def\url#1{\texttt{#1}}\fi
\expandafter\ifx\csname urlprefix\endcsname\relax\def\urlprefix{URL }\fi
\providecommand{\bibinfo}[2]{#2}
\providecommand{\eprint}[2][]{\url{#2}}

\bibitem[{\citenamefont{Dowson}(1979)}]{dowson}
\bibinfo{author}{\bibfnamefont{D.}~\bibnamefont{Dowson}},
  \emph{\bibinfo{title}{History of Tribology}} (\bibinfo{publisher}{Longman},
  \bibinfo{address}{New York}, \bibinfo{year}{1979}).

\bibitem[{\citenamefont{Baumberger and Caroli}(2006)}]{baumberger}
\bibinfo{author}{\bibfnamefont{T.}~\bibnamefont{Baumberger}} \bibnamefont{and}
  \bibinfo{author}{\bibfnamefont{C.}~\bibnamefont{Caroli}},
  \bibinfo{journal}{Adv. in Phys.} \textbf{\bibinfo{volume}{55}},
  \bibinfo{pages}{279} (\bibinfo{year}{2006}).

\bibitem[{\citenamefont{Capozza et~al.}(2009)\citenamefont{Capozza, Vanossi,
  Vezzani, and Zapperi}}]{capozzaprl}
\bibinfo{author}{\bibfnamefont{R.}~\bibnamefont{Capozza}},
  \bibinfo{author}{\bibfnamefont{A.}~\bibnamefont{Vanossi}},
  \bibinfo{author}{\bibfnamefont{A.}~\bibnamefont{Vezzani}}, \bibnamefont{and}
  \bibinfo{author}{\bibfnamefont{S.}~\bibnamefont{Zapperi}},
  \bibinfo{journal}{Phys. Rev. Lett.} \textbf{\bibinfo{volume}{103}},
  \bibinfo{pages}{085502} (\bibinfo{year}{2009}).

\bibitem[{\citenamefont{Vanossi et~al.}(2013)\citenamefont{Vanossi, Manini,
  Urbakh, Zapperi, and Tosatti}}]{rmpvanossi}
\bibinfo{author}{\bibfnamefont{A.}~\bibnamefont{Vanossi}},
  \bibinfo{author}{\bibfnamefont{N.}~\bibnamefont{Manini}},
  \bibinfo{author}{\bibfnamefont{M.}~\bibnamefont{Urbakh}},
  \bibinfo{author}{\bibfnamefont{S.}~\bibnamefont{Zapperi}}, \bibnamefont{and}
  \bibinfo{author}{\bibfnamefont{E.}~\bibnamefont{Tosatti}},
  \bibinfo{journal}{Rev. of Mod. Phys.} \textbf{\bibinfo{volume}{85}},
  \bibinfo{pages}{529} (\bibinfo{year}{2013}).

\bibitem[{\citenamefont{Capozza et~al.}(2013)\citenamefont{Capozza, Barel, and
  Urbakh}}]{capozza}
\bibinfo{author}{\bibfnamefont{R.}~\bibnamefont{Capozza}},
  \bibinfo{author}{\bibfnamefont{I.}~\bibnamefont{Barel}}, \bibnamefont{and}
  \bibinfo{author}{\bibfnamefont{M.}~\bibnamefont{Urbakh}},
  \bibinfo{journal}{Sci. Rep.} \textbf{\bibinfo{volume}{3}},
  \bibinfo{pages}{1896} (\bibinfo{year}{2013}).

\bibitem[{\citenamefont{Bowden and Tabor}(1950)}]{bowden}
\bibinfo{author}{\bibfnamefont{F.~P.} \bibnamefont{Bowden}} \bibnamefont{and}
  \bibinfo{author}{\bibfnamefont{D.}~\bibnamefont{Tabor}},
  \emph{\bibinfo{title}{The Friction and Lubrication of Solids}}
  (\bibinfo{publisher}{Oxford University Press}, \bibinfo{address}{New York},
  \bibinfo{year}{1950}).

\bibitem[{\citenamefont{Rubinstein et~al.}(2007)\citenamefont{Rubinstein,
  Cohen, and Fineberg}}]{fineb2007}
\bibinfo{author}{\bibfnamefont{S.~M.} \bibnamefont{Rubinstein}},
  \bibinfo{author}{\bibfnamefont{G.}~\bibnamefont{Cohen}}, \bibnamefont{and}
  \bibinfo{author}{\bibfnamefont{J.}~\bibnamefont{Fineberg}},
  \bibinfo{journal}{Phys. Rev. Lett.} \textbf{\bibinfo{volume}{98}},
  \bibinfo{pages}{226103} (\bibinfo{year}{2007}).

\bibitem[{\citenamefont{Ben-David et~al.}(2010)\citenamefont{Ben-David,
  Rubinstein, and Fineberg}}]{fineb2010}
\bibinfo{author}{\bibfnamefont{O.}~\bibnamefont{Ben-David}},
  \bibinfo{author}{\bibfnamefont{S.~M.} \bibnamefont{Rubinstein}},
  \bibnamefont{and} \bibinfo{author}{\bibfnamefont{J.}~\bibnamefont{Fineberg}},
  \bibinfo{journal}{Nature} \textbf{\bibinfo{volume}{76}}, \bibinfo{pages}{463}
  (\bibinfo{year}{2010}).

\bibitem[{\citenamefont{Otsuki and Matsukawa}(2013)}]{SRotsuki2013}
\bibinfo{author}{\bibfnamefont{M.}~\bibnamefont{Otsuki}} \bibnamefont{and}
  \bibinfo{author}{\bibfnamefont{H.}~\bibnamefont{Matsukawa}},
  \bibinfo{journal}{Sci. Rep.} \textbf{\bibinfo{volume}{3}},
  \bibinfo{pages}{1586} (\bibinfo{year}{2013}).

\bibitem[{\citenamefont{Ben-David and Fineberg}(2011)}]{fineb2011}
\bibinfo{author}{\bibfnamefont{O.}~\bibnamefont{Ben-David}} \bibnamefont{and}
  \bibinfo{author}{\bibfnamefont{J.}~\bibnamefont{Fineberg}},
  \bibinfo{journal}{Phys. Rev. Lett.} \textbf{\bibinfo{volume}{106}},
  \bibinfo{pages}{254301} (\bibinfo{year}{2011}).

\bibitem[{\citenamefont{Maegawa et~al.}(2010)\citenamefont{Maegawa, Suzuki, and
  Nakano}}]{Maegawa}
\bibinfo{author}{\bibfnamefont{S.}~\bibnamefont{Maegawa}},
  \bibinfo{author}{\bibfnamefont{A.}~\bibnamefont{Suzuki}}, \bibnamefont{and}
  \bibinfo{author}{\bibfnamefont{K.}~\bibnamefont{Nakano}},
  \bibinfo{journal}{Tribol. Lett.} \textbf{\bibinfo{volume}{38}},
  \bibinfo{pages}{313} (\bibinfo{year}{2010}).

\bibitem[{\citenamefont{Braun et~al.}(2009)\citenamefont{Braun, Barel, and
  Urbakh}}]{Braun}
\bibinfo{author}{\bibfnamefont{O.~M.} \bibnamefont{Braun}},
  \bibinfo{author}{\bibfnamefont{I.}~\bibnamefont{Barel}}, \bibnamefont{and}
  \bibinfo{author}{\bibfnamefont{M.}~\bibnamefont{Urbakh}},
  \bibinfo{journal}{Phys. Rev. Lett.} \textbf{\bibinfo{volume}{103}},
  \bibinfo{pages}{194301} (\bibinfo{year}{2009}).

\bibitem[{\citenamefont{Amundsen et~al.}(2012)\citenamefont{Amundsen,
  Scheibert, Thogersen, Tromborg, and Malthe-Sorenssen}}]{tromborg2012}
\bibinfo{author}{\bibfnamefont{D.~S.} \bibnamefont{Amundsen}},
  \bibinfo{author}{\bibfnamefont{J.}~\bibnamefont{Scheibert}},
  \bibinfo{author}{\bibfnamefont{K.}~\bibnamefont{Thogersen}},
  \bibinfo{author}{\bibfnamefont{J.}~\bibnamefont{Tromborg}}, \bibnamefont{and}
  \bibinfo{author}{\bibfnamefont{A.}~\bibnamefont{Malthe-Sorenssen}},
  \bibinfo{journal}{Tribol. Lett.} \textbf{\bibinfo{volume}{45}},
  \bibinfo{pages}{357} (\bibinfo{year}{2012}).

\bibitem[{\citenamefont{Tromborg et~al.}(2011)\citenamefont{Tromborg,
  Scheibert, Amundsen, Thogersen, and Malthe-Sorenssen}}]{tromborg2011}
\bibinfo{author}{\bibfnamefont{J.}~\bibnamefont{Tromborg}},
  \bibinfo{author}{\bibfnamefont{J.}~\bibnamefont{Scheibert}},
  \bibinfo{author}{\bibfnamefont{D.~S.} \bibnamefont{Amundsen}},
  \bibinfo{author}{\bibfnamefont{K.}~\bibnamefont{Thogersen}},
  \bibnamefont{and}
  \bibinfo{author}{\bibfnamefont{A.}~\bibnamefont{Malthe-Sorenssen}},
  \bibinfo{journal}{Phys. Rev. Lett.} \textbf{\bibinfo{volume}{107}},
  \bibinfo{pages}{07431} (\bibinfo{year}{2011}).

\bibitem[{\citenamefont{Burridge and Knopoff}(1967)}]{bk1}
\bibinfo{author}{\bibfnamefont{R.}~\bibnamefont{Burridge}} \bibnamefont{and}
  \bibinfo{author}{\bibfnamefont{L.}~\bibnamefont{Knopoff}},
  \bibinfo{journal}{Bull. Seismol. Soc. Am.} \textbf{\bibinfo{volume}{57}},
  \bibinfo{pages}{341} (\bibinfo{year}{1967}).

\bibitem[{\citenamefont{Carlson and Langer}(1989)}]{carlson}
\bibinfo{author}{\bibfnamefont{J.}~\bibnamefont{Carlson}} \bibnamefont{and}
  \bibinfo{author}{\bibfnamefont{J.}~\bibnamefont{Langer}},
  \bibinfo{journal}{Phys. Rev. Lett.} \textbf{\bibinfo{volume}{62}},
  \bibinfo{pages}{2632} (\bibinfo{year}{1989}).

\bibitem[{\citenamefont{Mori and Kawamura}(2008{\natexlab{a}})}]{bk2D}
\bibinfo{author}{\bibfnamefont{T.}~\bibnamefont{Mori}} \bibnamefont{and}
  \bibinfo{author}{\bibfnamefont{H.}~\bibnamefont{Kawamura}},
  \bibinfo{journal}{J. Geophys. Res.} \textbf{\bibinfo{volume}{113}},
  \bibinfo{pages}{B06301} (\bibinfo{year}{2008}{\natexlab{a}}).

\bibitem[{\citenamefont{Mori and Kawamura}(2008{\natexlab{b}})}]{bk2Dlr}
\bibinfo{author}{\bibfnamefont{T.}~\bibnamefont{Mori}} \bibnamefont{and}
  \bibinfo{author}{\bibfnamefont{H.}~\bibnamefont{Kawamura}},
  \bibinfo{journal}{Phys. Rev. E} \textbf{\bibinfo{volume}{77}},
  \bibinfo{pages}{051123} (\bibinfo{year}{2008}{\natexlab{b}}).

\bibitem[{\citenamefont{Ciamarra et~al.}(2010)\citenamefont{Ciamarra,
  Lippiello, Godano, and de~Arcangelis}}]{prl2010}
\bibinfo{author}{\bibfnamefont{M.~P.} \bibnamefont{Ciamarra}},
  \bibinfo{author}{\bibfnamefont{E.}~\bibnamefont{Lippiello}},
  \bibinfo{author}{\bibfnamefont{C.}~\bibnamefont{Godano}}, \bibnamefont{and}
  \bibinfo{author}{\bibfnamefont{L.}~\bibnamefont{de~Arcangelis}},
  \bibinfo{journal}{Phys. Rev. Lett.} \textbf{\bibinfo{volume}{104}},
  \bibinfo{pages}{238001} (\bibinfo{year}{2010}).

\bibitem[{\citenamefont{Ciamarra et~al.}(2011)\citenamefont{Ciamarra,
  Lippiello, de~Arcangelis, and Godano}}]{epl2011}
\bibinfo{author}{\bibfnamefont{M.~P.} \bibnamefont{Ciamarra}},
  \bibinfo{author}{\bibfnamefont{E.}~\bibnamefont{Lippiello}},
  \bibinfo{author}{\bibfnamefont{L.}~\bibnamefont{de~Arcangelis}},
  \bibnamefont{and} \bibinfo{author}{\bibfnamefont{C.}~\bibnamefont{Godano}},
  \bibinfo{journal}{Europhys. Lett.} \textbf{\bibinfo{volume}{95}},
  \bibinfo{pages}{54002} (\bibinfo{year}{2011}).

\bibitem[{\citenamefont{Nelson and Halperin}(1979)}]{nelson}
\bibinfo{author}{\bibfnamefont{D.~R.} \bibnamefont{Nelson}} \bibnamefont{and}
  \bibinfo{author}{\bibfnamefont{B.~I.} \bibnamefont{Halperin}},
  \bibinfo{journal}{Phys. Rev. B} \textbf{\bibinfo{volume}{19}},
  \bibinfo{pages}{2457} (\bibinfo{year}{1979}).

\bibitem[{\citenamefont{Silbert et~al.}(2001)\citenamefont{Silbert, Ertas,
  Grest, Halsey, Levine, and Plimpton}}]{silbert}
\bibinfo{author}{\bibfnamefont{L.~E.} \bibnamefont{Silbert}},
  \bibinfo{author}{\bibfnamefont{D.}~\bibnamefont{Ertas}},
  \bibinfo{author}{\bibfnamefont{G.~S.} \bibnamefont{Grest}},
  \bibinfo{author}{\bibfnamefont{T.~C.} \bibnamefont{Halsey}},
  \bibinfo{author}{\bibfnamefont{D.}~\bibnamefont{Levine}}, \bibnamefont{and}
  \bibinfo{author}{\bibfnamefont{S.~J.} \bibnamefont{Plimpton}},
  \bibinfo{journal}{Phys. Rev. E} \textbf{\bibinfo{volume}{64}},
  \bibinfo{pages}{051302} (\bibinfo{year}{2001}).

\end{thebibliography}

\end{document}